%% file: knactor.tex
\documentclass{acm}

\usepackage{times}  
\usepackage{array, graphics, graphicx, color, xspace, url, multirow, rotating, epsfig, amsmath, subfig}
\usepackage{wrapfig}
\usepackage{tikz}
\usepackage{booktabs}
\usepackage{epstopdf,subfig}
\usepackage{multicol}
\usepackage[format=plain,labelfont=bf,font=small]{caption}
\usepackage{ragged2e,flushend,enumitem,listings}
\usepackage{acronym} 
\usepackage{breakurl}
\usepackage{hyperref, cleveref}
\usepackage{cite}
\usepackage{multirow}
\usepackage[frozencache,cachedir=.]{minted}
\usepackage{appendix}
\usepackage[compact,small]{titlesec}
\usepackage[para]{footmisc}

\hypersetup{pdfstartview=FitH,pdfpagelayout=SinglePage}
\usepackage{enumitem}
\setenumerate[1]{label=\arabic*.}
\setenumerate[2]{label=\alph*.}
\setenumerate[3]{label=\roman*.}

\crefformat{section}{\S#2#1#3}
\crefformat{subsection}{\S#2#1#3}
\crefformat{subsubsection}{\S#2#1#3}

\setminted{fontsize=\footnotesize, xleftmargin=0.22in,linenos}

\newcommand{\bi}{\begin{itemize}}
\newcommand{\ei}{\end{itemize}}

\newcommand{\ie}{\emph{i.e.,}\xspace}
\newcommand{\eg}{\emph{e.g.,}\xspace}

\newcommand\paragraphb[1]{\noindent{\bf{#1}}}
\newcommand\paragraphi[1]{\noindent\emph{#1}}
\newcommand\pb[1]{\paragraphb{#1}}
\newcommand\pghi[1]{\paragraphi{#1}}

\newcommand{\figspace}{\vspace{-10pt}}

\newcommand{\eat}[1]{}

\newcommand{\Dex}{Data Exchange\xspace} 
 
\newcommand{\dex}{data exchange\xspace}

\newcommand{\DX}{DE\xspace} 
\newcommand{\DXs}{DEs\xspace}

\hypersetup{                    %
  colorlinks,
  linkcolor={green!80!black},
  citecolor={red!70!black},
  urlcolor={blue!70!black}
}

\definecolor{codegray}{rgb}{0.5,0.5,0.5}
\lstdefinestyle{bashStyle}{
    commentstyle=\color{blue},
    numberstyle=\tiny\color{codegray},
    basicstyle=\ttfamily\scriptsize,
    breakatwhitespace=false,         
    breaklines=true,                 
    captionpos=b,                    
    keepspaces=true,                 
    numbers=left,                    
    numbersep=5pt,                  
    showspaces=false,                
    showstringspaces=false,
    showtabs=false,                  
    tabsize=2,
    keywordstyle=\bfseries,
}

\graphicspath{{./dia/}}
\begin{document}

\title{From Kubernetes to Knactor: \\A Data-Centric Rethink of Service Composition} 
\author{Silvery Fu$^1$, Hong Zhang$^2$, Ryan Teoh$^1$, Taras Priadka$^1$, Sylvia Ratnasamy$^1$ \\ $^1$UC Berkeley, $^2$University of Waterloo}
\maketitle

\begin{abstract}
Microservices are increasingly used in modern applications, leading to a growing need for effective service composition solutions. However, we argue that traditional \emph{API-centric composition} mechanisms (\eg RPC, REST, and Pub/Sub) hamper the modularity of microservices. These mechanisms introduce rigid code-level coupling, scatter composition logic, and hinder visibility into cross-service data exchanges. Ultimately, these limitations complicate the maintenance and evolution of microservice-based applications. In response, we propose a rethinking of service composition and present Knactor, a new \emph{data-centric} composition framework to restore the modularity that microservices were intended to offer. Knactor decouples service composition from service development, allowing composition to be implemented as explicit data exchanges among multiple services. Our initial case study suggests that Knactor simplifies service composition and creates new opportunities for optimizations.
\end{abstract}

\input{intro}

\input{motif}

\input{design}

\input{proto}
\input{analysis}

\input{future}

\onecolumn \begin{multicols}{2}
\titlespacing*{\section}{0pt}{0pt}{-20pt}
\bibliographystyle{abbrv} 
\bibliography{knactor}
\end{multicols}

\end{document}

%% file: intro.tex
\section{Introduction}
\label{sec:intro}

The microservice architecture has been widely adopted in building modern applications~\cite{statista-microservice,alliedmarketresearch}. By breaking software systems down into smaller, independently deployable services, microservices make it easier to distribute development across teams and organizations such as SaaS providers and open-source communities. As more applications embrace this architecture, such as web and mobile back-ends, datacenter management, robot coordination, and IoT/smart space controllers~\cite{spring,asplos19-deathstar,sosp21-dspace,k8s,azure-drone,gcp-boutique}, the need for effective \emph{service composition} becomes increasingly important. Service composition combines multiple microservices to create an \emph{end-to-end application (app)}, making it a crucial and challenging aspect in application development. 

\begin{figure}[t]
     \centering
     \footnotesize
     \includegraphics[width = 0.48\textwidth]{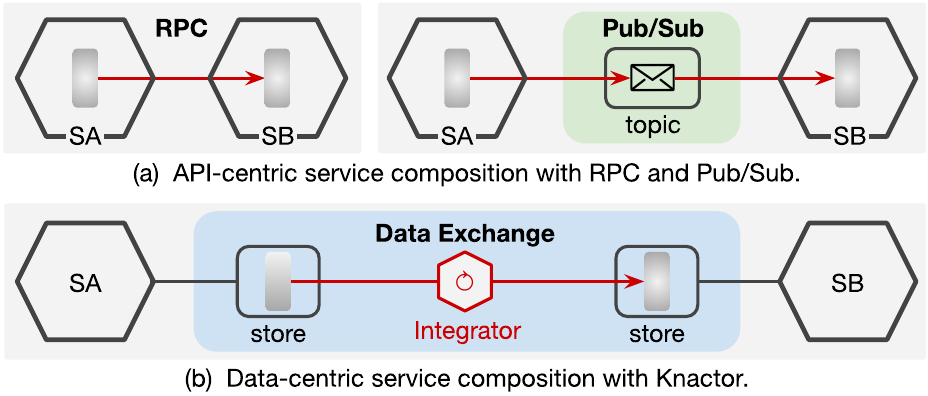}
     \figspace
     \caption{Comparison of service composition mechanisms. \rm{In RPC, Service $S_A$ invokes the API defined by Service $S_B$. In Pub/Sub, $S_A$ publishes messages to a topic where $S_B$ subscribes and receives these messages. In Knactor, $S_A$ and $S_B$ externalize their states in data stores hosted on a \dex. An integrator processes and syncs states between the data stores.}}
     \label{fig:mechanism}
     \figspace
\end{figure}

Today, service composition is typically achieved via remote procedure calls (RPCs)~\cite{grpc} or via publish-subscribe messaging (Pub/Sub)~\cite{kafka}, as depicted in Fig.\ref{fig:mechanism}a. With RPCs, service $S_A$ invokes the API of another service $S_B$ by incorporating the API endpoints and message schemas defined by $S_B$ (\eg gRPC and Protobuf~\cite{grpc}). Then, at run-time, $S_A$ sends a request message via a synchronous call to $S_B$ and, after running its procedure, $S_B$ replies with a response message. Pub/Sub replaces synchronous communication with asynchronous delivery of messages. In Pub/Sub, $S_B$ subscribes to a topic on a message broker (\eg Kafka~\cite{kafka}). $S_A$ can then send messages to this topic, which $S_B$ receives asynchronously and decodes using a schema. In this case, the topic and the schema (predefined by either $S_A$ or $S_B$) can be viewed as a variant of an API endpoint. We refer to these composition mechanisms as \emph{API-centric composition} where services are composed using predefined APIs at the time of development.

However, the API-centric approach leads to three drawbacks for service composition. First, it introduces tight \emph{coupling} between services. To compose services $S_B$ to $S_A$, service developers incorporate $S_B$'s schemas, client code stubs, invocation methods, response and error handling in the $S_A$'s code itself. Consequently, making composition changes, such as replacing $S_B$ with $S_C$ or adapting $S_A$ to new schema of $S_B$, requires accessing and modifying the code of $S_A$ as well as rebuilding and redeploying it. 
Second, in an application consisting of $N$ services, the composition logic is scattered across $O(N)$ services. Each service may use $O(N)$ external APIs, while its own APIs are used by $O(N)$ other services, depending on the in/out-degree~\cite{luo2021characterizing}. As a result, changes in the composition logic often involve and impact many services in the application, requiring extensive coordination and code-level changes across developers and teams responsible for each service. Third, the composition logic---especially the \emph{data exchanges} among services---are \emph{hidden} within the API invocations between individual pairs of services. This lack of visibility hinders customizing and optimizing composition at the app-level and at run-time. 

The above drawbacks suggest that \emph{while microservices are modular, existing composition mechanisms hamper this modularity}, complicating both the development and the composition of microservices. In particular, as modern applications increasingly consist of 10s, 100s, or even 1,000s of microservices (\eg as seen in web apps~\cite{airbnb,netflix,uber}), implementing and evolving service composition that is coupled, scattered, and hidden among individual services can become painstaking and prone to errors (\S\ref{sec:motif}). Such complexity escalates when apps and services are developed by different companies or vendors (\eg as seen in IoT apps~\cite{iot-analytics,smartthings} and public API marketplaces~\cite{rapidapi}) due to the high communication and coordination costs for making service changes~\cite{microservice}.

How can we simplify service composition and restore the modularity that microservices were designed to offer? In this paper, we propose two guiding principles for service composition: \textbf{(P1)} \emph{Decouple service composition from service development}, enabling the flexibility to implement and consolidate service composition after the development phase; and \textbf{(P2)} \emph{Make data exchanges explicit}, providing greater visibility and control to simplify service composition.

\begin{figure}
     \centering
     \footnotesize
     \includegraphics[width = 0.48\textwidth]{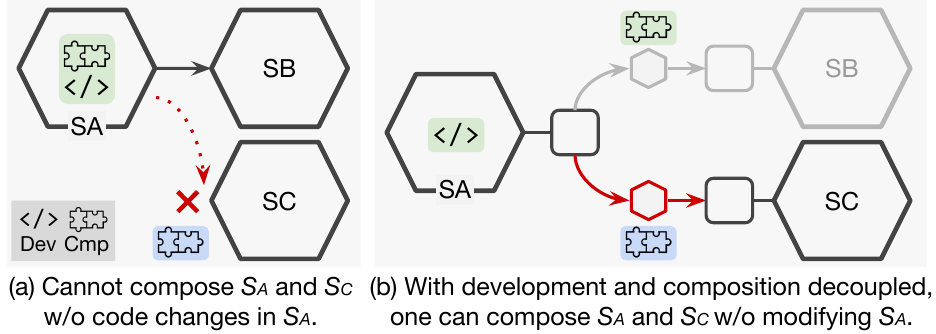}
     \figspace
     \caption{Decoupling service development and composition. \rm{Existing composition mechanisms, shown in (a), couple the service development and composition. Knactor, shown in (b), allows these two to be decoupled. \textbf{Dev:} Service development; \textbf{Int:} Service composition.}}
     \label{fig:role}
     \figspace
\end{figure}

We present a new service composition mechanism that follows principles P1 and P2. Our key insight is that services should be composed \emph{via states, not APIs}: we propose replacing API-centric composition with a new approach that we term \emph{data-centric} composition. In this data-centric approach, there are no RPCs or messaging between services, \ie no direct invocation of each others' APIs. Instead, each service externalizes its states to an associated \emph{data store} (Fig.\ref{fig:mechanism}b),\footnote{Note that unlike in Pub/Sub, in our model, a service does not subscribe or publish to a different service but only to its \emph{own} data store.} and an \emph{integrator} module acts as the intermediary that composes services by processing and syncing states between their data stores. The integrator can be easily replaced or reconfigured, not only during development but also at run-time (P1). In the integrator, developers can use dedicated state processing primitives, such as data exchange graphs and dataflow operators (\S\ref{sec:design}), to conveniently specify the desired data exchanges among services (P2).

We call this approach to developing and integrating services as the ``Knactor pattern.''\footnote{Knactor (pronounced ``connector'') stands for Kubernetes-native actors in a nod to Kubernetes~\cite{k8s} and the actor model~\cite{actor,actor-gul} from which we drew inspiration (see \S\ref{sec:future}).} The Knactor pattern addresses the drawbacks of existing service composition mechanisms through its core design choices: \textbf{(i)} Knactor confines the interaction of each service to its own data store as opposed to introducing code-level coupling to other services. The composition logic is implemented in the integrator without modifying the services' code (Fig.\ref{fig:role}) and can be reconfigured even at run-time. \textbf{(ii)} Knactor consolidates the composition logic into a single or a few application-level integrator modules, as opposed to the $O(N)$ services and their codebases; thus making it easier to implement composition logic, reducing excessive coordination and communication across teams or companies. \textbf{(iii)} By having services externalize their states, developers gain the ability to implement composition in the form of data exchanges, with the help of dedicated state processing primitives; as opposed to dealing with intricate sequences of API invocations across different services. This approach also makes the composition code easier to change and maintain.

We further discuss the motivation for the Knactor pattern (\S\ref{sec:motif}) and propose our design for the \emph{Knactor framework} to support this pattern in microservices (\S\ref{sec:design}). We report on our early experiences with prototyping the Knactor framework (\S\ref{sec:proto}) and using it in example Web and IoT apps (\S\ref{sec:analysis}). Our initial case study indicates that Knactor can greatly simplify service composition, reducing both development and operational efforts. While the modularity introduced by Knactor does bring some overhead, its overall performance impact on the applications we studied is minimal. Importantly, this modularity also opens up opportunities for performance optimizations, although for highly latency-sensitive applications direct RPCs may still be preferable. We discuss related work and outline future research directions with Knactor (\S\ref{sec:future}).

%% file: motif.tex
\section{Background and Motivation}
\label{sec:motif} 
While microservices, and more broadly the \emph{service-oriented architecture} (SOA~\cite{soa}), offer flexibility and scalability, they also create the need for effective composition and cooperation between services and teams. Today, composition is based on communication protocols such as RPC~\cite{grpc,thrift}, message brokers~\cite{kafka,emqx}, and REST APIs~\cite{rest,graphql}. In what follows, we use examples from real-world apps to review the development workflow with existing composition mechanisms and motivate the case for a new approach. We'll use these apps in \S\ref{sec:design}-\S\ref{sec:analysis} and show how they're handled with Knactor.

\begin{figure}[t]
     \centering
     \footnotesize
     \includegraphics[width = 0.48\textwidth]{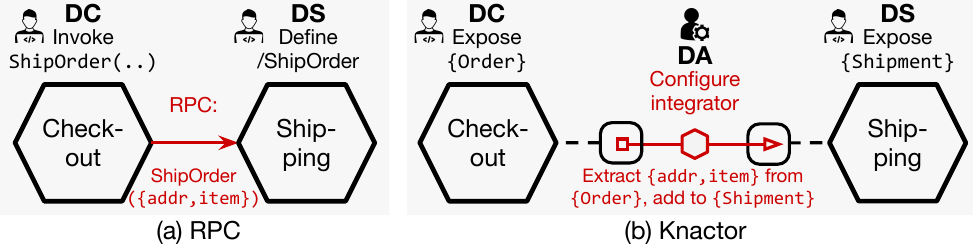}
     \caption{Comparison of composition mechanisms in an online retail app.}
     \label{fig:retail}
     \figspace
\end{figure}

\pb{(1) Web: Online Retail.} The first example is part of an online retail web application~\cite{gcp-boutique}. The original app uses gRPC to compose services~\cite{grpc}. Our focus is on two microservices, Checkout ($C$) and Shipping ($S$), and on the shipment request from $C$ to $S$, which creates shipments for orders that have been checked out. With RPC, the developer of $S$ ($D_S$) defines API endpoints with the API's name, version, and the message schema(s), etc. for the request messages in a Protobuf~\cite{protobuf} definition file. As shown in Fig.\ref{fig:retail}a, in this example, the API endpoint is \texttt{/ShipOrder} which takes item name and shipment address as its inputs. Then, $D_S$ may share the API definition file with the developer of $C$ ($D_C$), who uses the file to create the client stub code, imports and uses the code in $C$, and finally recompiles/builds $C$.

\pb{(2) IoT: Smart Home.} The second example is from a home automation app~\cite{smartthings,chi14-home-tap,sosp21-dspace} adapted from an open-source IoT app simulator~\cite{hotnets22-digibox}. The app includes a house service (developed by IoT company X, \eg Samsung SmartThings~\cite{smartthings}) that automatically adjusts the brightness level of lamps (from device vendor Y, \eg Lifx~\cite{lifx}) based on occupancy sensor readings (from device vendor Z, \eg Ring~\cite{ring}) while monitoring the energy consumption of these devices. There are three services in this application---House ($H$), Motion ($M$), and Lamp ($L$)---that are composed via the message broker EMQX~\cite{emqx}. $H$ subscribes to messages on $M$ (the ``motion'' topic), and when $H$ receives a message reporting ``triggered: true'', it publishes a message to $L$'s topic to adjust the lamp's brightness level. For each service, the developer uses Protobuf to define schemas for the messages exchanged among devices. For example, $H$ uses the schema of $M$ and $L$ to deserialize the messages from the two and vice versa.

\pb{Problem 1: Services are overly \emph{coupled}.} In both examples, we observe that API-centric composition introduces code-level coupling between services (\eg $C$ to $S$, $H$ to $L$ and $M$). Consequently, making any composition changes (\eg switching $C$ to a new shipping service) and ensuring compatibility with APIs (\eg adapting $C$ or $H$ to the changes of schema of $S$ or $L$) require modifications at the code level. This further leads to rebuilding and redeploying services, which also requires careful planning in the production environment to avoid application downtime~\cite{encora-zero-downtime,opslevel-deployment} and takes additional time and compute resources. For example, adapting $C$ to an API schema change in $S$ requires 69 lines of code and configuration updates (\S\ref{sec:analysis}), followed by recompiling $C$, updating and uploading its container images, and redeploying $C$ using a rolling update in Kubernetes~\cite{k8s-rolling-update}. Such changes are common in microservice-based applications~\cite{infoq-microservice, microsoft-versioning}.

\pb{Problem 2: Composition logic is \emph{scattered}.} A direct consequence of implementing service composition within individual services is the scattering of composition logic across multiple services and service codebases. In the web app we studied, we identified 15 methods on handling API invocations scattered across 11 services, and 36 across 14 services in another well-studied social networking app~\cite{social-network,deathstar}. It is worth noting that these examples represent only small-scale apps developed for demonstration purposes. In production, applications may have composition logic scattered across more than 100s/1,000s of microservices~\cite{uber,airbnb,luo2021characterizing}. 

\pb{Problem 3: Data exchanges are \emph{hidden}.} The examples also highlight that the data exchanges among services are hidden within pair-wise API invocations. In the online retail app, when an order checkout request is received by the $C$, it triggers an RPC request to the $S$. This request includes the order's states, which are not accessible outside this pair of services. Similarly, in the smart home system, state changes in the $M$ or the $L$ implicitly influence the state of the $H$ through the messages exchanged among these services. This lack of visibility hinders the ability to add functionalities (\eg implementing ``conditional composition'' where $C$ should opt for air shipping when the order's price exceeds $1,000$ USD) and access control (\eg $H$ should not access the $L$ during user-defined sleep hours).

%% file: design.tex
\section{Knactor Design}
\label{sec:design}
We present the rationale of the Knactor pattern (\S\ref{subsec:principle}), the framework's designs (\S\ref{subsec:framework}), and optimizations (\S\ref{subsec:opt}).

\subsection{Design Rationale}
\label{subsec:principle} 

Knactor's design is guided by two key principles aimed at enhancing modularity. The first is to \textbf{decouple service composition from service development}, which allows for the creation and updating of service composition outside the service development phase. This principle follows the classic design principle of \emph{separating mechanism and policy}, \ie the composition mechanism should not dictate which services \emph{can} be composed. The rationale is that at the development stage, it can be challenging to anticipate all the ways the service might be used and extended in apps (\ie the ``composition policy''). Thus, an overly constraining composition mechanism can substantially increase the cost and delay involved in policy changes post-development.

Knactor achieves a cleaner separation of composition mechanism and policy by enabling \emph{late-binding} of services via \emph{two levels of indirection}: a per-service data store and integrator. In Knactor (Fig.\ref{fig:mechanism}b), a service does not directly access other services' APIs nor their states but \emph{its own data store only}. Then, an integrator module acts as an intermediary, responsible for processing and syncing states across the data stores of the services being composed. 

The second guiding principle of Knactor is to replace API invocations with \textbf{explicit data exchanges} among services. The rationale is that developers, rather than calling and responding to APIs, can concisely specify data exchange patterns among multiple services and leverage the dedicated state processing primitives of a framework (\S\ref{subsec:framework}) for simplified composition. This approach also streamlines the maintenance and evolution of composition logic, as updating the composition logic only requires altering the data exchange specification, eliminating the need to rewire API calls (\S\ref{sec:analysis}). Finally, the declarative nature of data-centric composition separates the specification of composition from its execution, providing opportunities for optimization (\S\ref{subsec:opt}).

\subsection{Knactor Framework}
\label{subsec:framework}

The Knactor framework provides the programming libraries, tooling, and runtime to facilitate service composition with the data-centric approach. In Knactor, each microservice is represented as a \emph{knactor}\footnote{We overload the term ``Knactor'' to mean both the pattern and the framework, and the lowercase ``knactor'' the service abstraction.} that contains a \emph{reconciler} component and one or multiple \emph{data stores}. 

\begin{figure}[t]
     \centering
     \footnotesize
     \includegraphics[width = 0.47\textwidth]{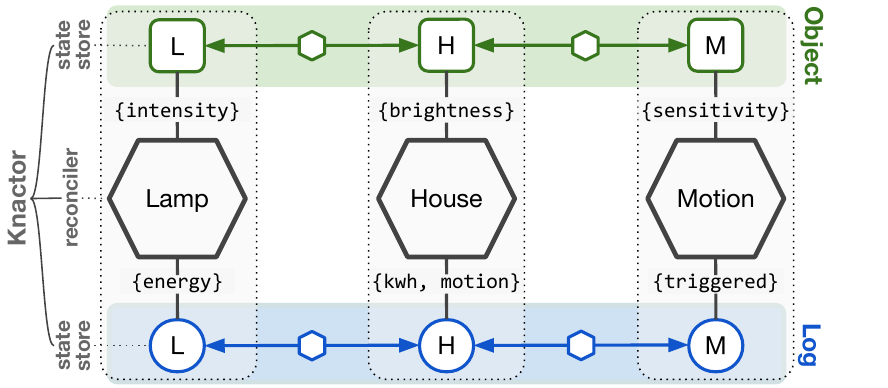}
     \figspace
     \caption{Building the smart home app in Knactor. \rm{There are three knactors, Lamp, House, and Motion each has two data stores, one on Object data exchange and one on Log data exchange.}}
     \label{fig:space}
     \figspace
\end{figure}

\pb{Data store and exchange.} A data store keeps the states relevant to the knactor's operation, such as order status in the Checkout service (Fig.\ref{code:checkout}). The data stores are hosted on a logically centralized \Dex (\DX) that provides state access and management capabilities such as data storage, caching, scaling, analytics, and access control. There can be different types of \DXs that each specialized at handling a different type of states/data, \eg API objects~\cite{k8s-apiserver}, logs~\cite{cidr23-zed}, and database tables~\cite{postgres}. We envision the \DXs will be taken off-the-shelf and the Knactor framework provides wrappers, tooling, and extensions around these \DXs for simplifying and optimizing composition (\S\ref{subsec:opt}). As a starting point, we focus on two types of \DXs, ``Object'' and ``Log''. The former keeps states as attribute-value pairs in a k-v store and exposes APIs for CRUD operations over these states, while the latter keeps states as structured and semi-structured data as append-only logs and exposes data ingestion and analytics APIs. A knactor can have multiple data stores and thus use multiple \DXs. For example, in Fig.\ref{fig:space}, the three knactors each have two data stores on Object and Log, with the ones on Object storing configuration states such as lamp's intensity level while the ones on Log storing sensor data such as motion readings.

\pb{Reconciler.} The reconciler is a code module that interacts with the knactor's data store(s) using the state access methods provided by the \DX. It responds to state updates from the data store and initiates corresponding actions. For example, the reconciler inside the Shipping knactor may process a new shipment object that appears in the data store (\eg by initiating a FedEx delivery~\cite{shipengine}), and can also update the data store, such as posting the shipment's tracking ID. Service developers can adapt existing services by removing any external API invocations they aim to decouple from the service and adding required state access to the data store, or they can design and implement a new reconciler from scratch.

\pb{Integrator.} An integrator syncs and processes states between data stores leveraging the APIs provided by the \DXs. For instance, the integrator in Fig.\ref{fig:retail}b can use the CRUD APIs provided by the Object \DX to obtain the order states from the Checkout service, extract the shipment states \texttt{items} and \texttt{addr}, and update the Shipping knactor's data store. The framework provides \emph{built-in integrators} specialized for processing states over a type of \DX and data exchange patterns. Developers can use the state processing primitives from a built-in integrator to implement composition (described next). We focus on two built-in integrators, Cast and Sync, that handle states on Object and Log respectively.

For example, as shown in Fig.\ref{fig:space}, the House knactor has a configuration for motion sensor's sensitivity in its data store, the Cast will post this to the Motion knactor's store. The Sync extracts data from the source store(s) on the Log and loads it to the destination store(s). Developers can write dataflow on the Cast and Sync's specifications to add additional data processing/ETL logic~\cite{etl}. Further, note that while Knactor allows for the consolidation of composition logic into a single or a few integrators, it \emph{does not impose} this as a requirement. Developers and service operators have the flexibility to implement integrators in ways that align with their organizational structure and preferences (\S\ref{sec:future}).

\begin{figure}
\begin{minted}{yaml}
schema: OnlineRetail/v1/Checkout/Order
items: object 
address: string
cost: number 
shippingCost: number # +kr: external
totalCost: number
currency: string
paymentID: string    # +kr: external
trackingID: string   # +kr: external
\end{minted}
\figspace 
\caption{Schema of the Checkout knactor's data store.}
\label{code:checkout}
\figspace
\end{figure}

\begin{figure}
\begin{minted}{yaml}
Input:
  C: OnlineRetail/v1/Checkout/knactor-checkout
  S: OnlineRetail/v1/Shipping/knactor-shipping
  P: OnlineRetail/v1/Payment/knactor-payment
DXG:
  C.order: 
    shippingCost: >
      currency_convert(S.quote.price,
                       S.quote.currency, this.currency)
    paymentID: P.id
    trackingID: S.id
  P:
    # other fields in the state store: id
    amount: C.order.totalCost
    currency: C.order.currency
  S: 
    # other fields in the state store: id, quote
    items: '[item.name for item in C.order.items]'
    addr: C.order.address
    method: > 
      "air" if C.order.cost > 1000 else "ground"
\end{minted}
\figspace
\caption{Specification of the data exchange graph (DXG) for the integrator in the online retail web app.}
\label{code:sdg}
\figspace
\figspace
\end{figure}

\pb{Development workflow.} The development workflow contains three logical steps: \textbf{(i) Externalize.} At development time, the developer registers the schema of the knactor's data store to the \DX. \textbf{(ii) Express.} The developer indicates what states its data store can ingest by annotating the fields in the data store, \eg in the Checkout knactor's data store (\eg Fig.\ref{code:checkout}) can indicate the ``shippingCost'', ``paymentID'', and ``trackingID'' are annotated to indicate they are to be filled externally by an integrator. Likewise, the House knactor can indicate it can ingest ``kwh'' and ``motion'' readings (Fig.\ref{fig:space}). \textbf{(iii) Exchange}. To compose services/knactors, the developer specifies the data exchanges among their data stores via programming or configuring the integrators. The state accesses and exchanges are subject to access control (\S\ref{subsec:opt}).

\pb{Data exchange primitives.} The integrator library provides primitives that simplify the expression of data exchange patterns for service composition. As shown in Fig.\ref{code:sdg}, the Cast integrator supports data exchange graphs (DXGs), allowing the declarative description of data exchanges among multiple services. It can include references to states in each service's data store, state transformation and aggregation functions, and data-centric policies. Similarly, the Sync integrator offers dataflow operators like filter, rename, sort, and aggregation functions. For example, the House can retrieve motion sensor readings from the Motion service, and the Sync integrator can rename the ``trigger'' field to ``motion'' before loading the data into the House's data store. 

\subsection{Run-time Operation and Optimization}
\label{subsec:opt}

\pb{Integrator reconfiguration.} Integrators, such as Cast and Sync, can be dynamically reconfigured at run-time to add new composition logic or modify existing configurations. This avoids service-level code changes, rebuilding, and redeployment for each composition update. For example, a new data-centric policy can easily be introduced to the DXG specification (Fig.\ref{code:sdg}, line 22) to determine the shipment method based on the order price, without changing Checkout or Shipping.

\pb{State retention.} By default, states in the data stores are preserved until they're no longer required by entities such as the knactor's reconciler or integrators. State retention can be managed via reference counting or similar mechanisms that track the usage of state objects.
Once a reconciler or integrator has performed its operation on a state object, the object is marked as unused and the \DXs can then perform garbage collection with standard recovery techniques. %
In addition, one can also specify customized state retention policies for archival or analytical purposes.

\pb{State access control.} Knactor ensures only authorized entities can access the states in the data stores. First, developers can only view data store schemas, not actual states. At run-time, access to a knactor's data store is limited to its own reconciler and any integrators that have been granted access through access control policies. This can be done via the standard Role-based Access Control (RBAC), assigning roles to reconcilers and integrators to manage access~\cite{k8s-access}. Second, the data-centric approach allows finer-grained access control over states~\cite{osdi22-blockaid}, \eg granting access to certain state objects/fields but not others to specific roles. 

\pb{Performance optimization.} Knactor can optimize data exchange performance and efficiency by leveraging the modularity of the data store and integrator. First, one can use \DXs optimized for high-performance such as in-memory k-v stores~\cite{redis}. The integrators can perform \emph{push-down optimization} to offload composition logic to the \DX using features such as user-defined functions (UDFs) and stored procedures~\cite{redis-udf} common in these \DXs. This can accelerate and reduce data movement between the \DX and integrator. Second, when data stores are hosted on the \DX, the \DX and integrator can implement \emph{zero-copy} data exchange to further minimize the data movement. Third, integrators can consolidate the state processing logic by combining multiple state processing operations into fewer and more efficient ones. For example, if multiple knactors are looking to consume data from the same Log data store and apply the same data transformations over raw data, Sync can optimize the dataflow operations between data stores to reuse intermediate results. We evaluate some of these optimizations in \S\ref{sec:analysis}.

%% file: proto.tex
\section{Prototype and Apps}
\label{sec:proto}

We built a prototype of the Knactor framework. It contains a programming library in Python for knactor development including communication packages for \DXs, code generators, and a CLI for operating knactors. For the two \DXs, we used open-source Kubernetes apiserver~\cite{k8s-apiserver} (and Redis~\cite{redis} as an alternative) for Object and Zed lake~\cite{cidr23-zed}) for Log; we implemented the built-in integrators, Cast and Sync, for Object and Log respectively. 

Using the Knactor prototype, we implemented the two example apps, online retail and smart home mentioned earlier. As shown in Fig.\ref{fig:retail}b, the online retail app consists of 11 knactors including the Checkout and Shipment discussed earlier, and a Cast-based integrator that composes the knactors. For the smart home app in Fig.\ref{fig:space}, we implemented House, Lamp~\cite{lifx}, and Motion~\cite{ring} knactors with device vendor libraries; as well as the Cast- and Sync-based integrators.

%% file: analysis.tex
\section{Preliminary Results}
\label{sec:analysis}

We studied how well Knactor simplifies service composition and its implications on performance using the prototype and apps. We report the preliminary results in what follows.

\begin{table}
\def\arraystretch{1.2}
\centering
\footnotesize
\begin{tabular}{|c|c|cc|cc|cc|}
\hline
\textbf{App} & \textbf{Task} & \multicolumn{2}{c|}{\textbf{Operation}} & \multicolumn{2}{c|}{\textbf{\# File}} & \multicolumn{2}{c|}{\textbf{SLOC}} \\ \hline
\multirow{4}{*}{\begin{tabular}[c]{@{}c@{}}Online\\ Retail\end{tabular}} & - & \multicolumn{1}{c|}{API} & \textbf{KN} & \multicolumn{1}{c|}{API} & \textbf{KN} & \multicolumn{1}{c|}{API} & \textbf{KN} \\ \cline{2-8} 
 & 1 & \multicolumn{1}{c|}{c / f / b / d} & \textbf{f} & \multicolumn{1}{c|}{8} & \textbf{1} & \multicolumn{1}{c|}{109} & \textbf{7} \\ \cline{2-8} 
 & 2 & \multicolumn{1}{c|}{c / f / b / d} & \textbf{f} & \multicolumn{1}{c|}{2} & \textbf{1} & \multicolumn{1}{c|}{14} & \textbf{1} \\ \cline{2-8} 
 & 3 & \multicolumn{1}{c|}{c / f / b / d} & \textbf{f} & \multicolumn{1}{c|}{4} & \textbf{1} & \multicolumn{1}{c|}{93} & \textbf{7} \\ \hline
\end{tabular}
\caption{{\bf Comparison of composition cost}: API-centric (API) vs. Knactor (KN). Annotations indicate required operations, \textbf{c:} code changes; \textbf{f:} configuration changes. \textbf{b:} rebuild service; \textbf{d:} redeploy service.}
\label{tab:cost}
\figspace
\end{table}

\pb{Composition cost.} We compare the service composition effort required for the Knactor and the API-centric approach using various tasks in the online retail app. The tasks include $T_1$: integrating the Payment and Shipping services with the Checkout service; $T_2$: adding a shipment policy based on the order price; and $T_3$: updating the Shipping schema. We compare the required operations (\eg code modifications, configuration updates, build and deployment steps), number of files, and the source lines of code (SLOC) changed or used to implement the task, including the services' source code, scripts, configurations, and schema definitions. 

\pghi{\underline{Takeaways.}} Table~\ref{tab:cost} presents the results. As shown, (i) In all tasks $T_1$-$T_3$, composition changes using Knactor require only reconfiguring the integrator module, while the API-centric approach requires code and configuration changes as well as service rebuilds and redeployments. This demonstrates Knactor's benefit of decoupling composition from development through externalized states and integrator, which facilitates composition without code-level changes at the individual services. (ii) Knactor enables the consolidation of composition logic, allowing modifications to be made in a single location (\eg the DXG configuration) instead of multiple files across separate services/service codebases, which further simplifies the composition task. (iii) While the core composition logic remains the same in both approaches, compared to the API-centric approach, Knactor simplifies the \emph{implementation} of composition logic, reducing the required SLOC (\eg by 102 in $T_1$). We conjecture this is due to the \emph{imperative} nature of the API-centric approach towards composition, which incurs not only the overhead of handling the ``mechanics'' (\eg importing and handling schemas and client stubs), but also the complexity of expressing interactions across services (\eg Checkout, Shipping, and Payment) as a sequence of API invocations. In contrast, Knactor enables the composition task to be \emph{declaratively} and \emph{concisely} expressed as data exchanges over externalized states (\eg DXG in Fig.\ref{code:sdg}), which saves the overhead on the mechanics of composition and inherently captures the ordering information as state dependencies get resolved. 

\pb{Impact on application performance.} Compared to RPC, Knactor introduces two indirections for modularity: the data store and the integrator. To understand their impact on application performance, we benchmark the Cast between the Checkout and Shipping knactors deployed on a Kubernetes cluster. We measure the state propagation latency, with and without optimizations (\S\ref{subsec:opt}), and repeat for the API-centric baselines and compare the results.

\begin{table}
\centering
\footnotesize
\begin{tabular}{|c|c|c|c|c|c|c|}
\hline
\textbf{Setup} & \textbf{C-I} & \textbf{I} & \textbf{I-S} & \textbf{S} & \textbf{Prop} & \textbf{Total (ms)} \\ \hline
RPC & - & - & - & 446 & \textbf{1.8} & \textbf{447.8} \\ \hline
K-apiserver & 20.6 & 0.01 & 12.5 & 453 & 33.1 & 486.1 \\ \hline
K-redis & 3.2 & 0.06 & 2.7 & 444 & 5.8 & 449.8 \\ \hline
K-redis-udf & 2.1 & 0.7 & 0.1 & 450 & \textbf{2.9} & \textbf{452.9} \\ \hline
\end{tabular}
\caption{Latency in the online retail app completing a shipment request, \rm{with breakdown by stage. \textbf{C-I}: Checkout and integrator. \textbf{I}: Integrator. \textbf{I-S}: Integrator and Shipping. \textbf{S:} Shipment processing.} \textbf{Prop:} State propagation.}
\label{tab:prop}
\figspace  
\end{table}
\pghi{\underline{Takeaways.}} Table~\ref{tab:prop} provides a breakdown of latencies in the online retail application. We compare three Knactor configurations: K-apiserver (using a strongly consistent k-v store with persistent storage~\cite{etcd} for Object exchange), K-redis (employing Redis, an in-memory data store~\cite{redis}), and K-redis-udf (incorporating an integrator pushdown optimization using Redis's UDF~\cite{redis-udf}). First, the choice of \DX substantially impacts the state propagation latency ($33.1 ms$ in K-apiserver vs. $5.8 ms$ in K-redis); a high-performance \DX can significantly reduce the data movement overhead from the integrator and reconciler. This overhead can be further reduced via integrator pushdown (\eg from $2.7 ms$ to $0.1 ms$ between the integrator and Shipping's data store, with K-redis-udf). Second, the overhead has a relatively small impact on the performance of this app we studied, with the Shipment processing~\cite{shipengine} as the primary bottleneck. However, note that Knactor does lead to higher latency overhead for state propagation; thus, for highly latency-sensitive workloads~\cite{nsdi19-shenango}, direct RPC may still be the preferred approach. 

Moreover, our benchmark on the IoT app and Sync---details omitted due to space---indicates that the consolidated state processing optimization (\S\ref{subsec:opt}) can reduce the average state processing latency by up to 3.5x. This is achieved by minimizing redundant operations that apply the same data transformations over sensor knactors' states. Comprehensive evaluation on performance and scalability is our future work.

%% file: future.tex
\section{Discussion}
\label{sec:future}
\vspace{0.5em}
In this section, we expand on related work beyond what was mentioned in \S\ref{sec:motif}, followed by future directions in Knactor.

\pb{Kubernetes and actor model.} Knactor's design is influenced by Kubernetes~\cite{cacm16-bok}, a cluster management system that manages applications via controllers~\cite{k8s-controller}. The controllers read from and write to shared API/resource objects on an API server~\cite{k8s-apiserver} as a means to interact. Compared to Knactor, Kubernetes does not contain the integrator abstraction, separation of data stores, and specialized data exchange designs. %
Knactor is also inspired by the actor model~\cite{actor,actor-gul}, where actors are separate units of computation that interact via message passing. Knactor can be seen as consisting of two types of actors, knactors that express the main service/business logic and integrators that implement the composition logic and data exchanges between knactors. 

\pb{Apps and applicability.} Knactor is particularly beneficial for applications with many microservices and complex compositions, such as cellular EPC~\cite{nsdi23-magma} besides the ones discussed in this paper. Currently, these systems rely on many inter-service APIs which leads to significant complexity~\cite{sigcomm21-cellbricks} and is difficult to evolve. Our ongoing work focuses on applying Knactor to different application domains, from large-scale web applications~\cite{deathstar}, and IoT systems~\cite{sosp21-dspace}, to datacenter and EPC systems~\cite{nsdi23-magma} to understand the benefits.

Knactor also offers the benefits of enabling service composition to be performed by individuals who are not the original service developers. Similar to how API definitions and documentation convey information about the behaviors and semantics of today's services, Knactor developers can use the data store schema and other documentation to obtain required composition-related information. If these resources do not sufficiently describe the services' behaviors for effective composition, one can still fall back on engaging the original service developers to implement the composition task.

\pb{Framework support for composition.} The visibility over states and data exchanges in Knactor allows developers to leverage tools such as formal methods and static analysis~\cite{panda2017verification} as well as run-time primitives such as transactions~\cite{cheng2021ramp} for implementing composition at large-scale. For example, the Cast can provide loop and unused state detection with static analysis to assist developers build robust data exchanges.

\pb{Compatibility and deployability.} We expect the use of Knactor with existing systems can be facilitated through the use of proxies or porting mechanisms~\cite{envoy,nsdi23-mrpc}. Deployment issues such as load balancing, autoscaling, and observability, such as monitoring knactor SLOs through distributed tracing~\cite{sigelman2010dapper,hotnets21-snicket} and telemetry~\cite{leffler2022opentelemetry}, are also worth exploring.

\pb{Ecosystem.} Knactor could potentially have far-reaching implications for the service and application ecosystem. For example, a marketplace for knactors and integrators could emerge, akin to current API marketplaces~\cite{rapidapi}. In such a marketplace, knactors and integrators, developed by various individuals or organizations, could be shared and reused, fostering more collaborative and streamlined app development.